\documentclass[11pt,twoside]{article}
\usepackage{amsmath,graphicx}

\begin{document}

\title{Effects of Cavity and Superradiance on the Electrical Transport
Through Quantum Dots}
\author{Y.N. Chen$^{1}$, D.S. Chuu$^{1}$ and T. Brandes$^{2}$}
\date{}
\maketitle

\noindent $^{1}$Department of Electrophysics, National Chiao-Tung
University,
Hsinchu 300, Taiwan\\
$^{2}$Department of Physics, UMIST, P.O. Box 88, Manchester, M60
1QD, U.K.

\begin{abstract}
A novel method is proposed to measure the Purcell effect by observing the
current through a semiconductor quantum dot embedded inside a microcavity.
The stationary current is shown to be altered if one varies the cavity
length. For the double-dot system, the stationary current is found to show
the interference feature (superradiance) as the inter-dot distance is
varied. The amplitude of oscillation can be increased by incorporating the
system into a microcavity. Furthermore, the current is suppressed if the dot
distance is small compared to the wavelength of the emitted photon. This
photon trapping phenomenon generates the entangled state and may be used to
control the emission of single photons at predetermined times.
\end{abstract}

\section{Introduction}

\ \ \ \ \ Since Dicke proposed the idea of superradiance\cite{1}, coherent
radiation phenomena for atomic systems was intensively investigated\cite%
{2,3,4,5,6}. What Dicke found was that when the gas is in a particular state
with the half number of molecules excited, the spontaneous emission rate of
the whole gas is proportional to the square of the molecular concentration,
provided that the gas volume dimension is small compared to the emitted
photon wavelength. One of the interests in superradiant study lies in its
close connection with the physics of laser emission. In some aspect, the
superradiant phenomena appears somewhat simpler since one can neglect the
pumping and relaxation mechanisms which are important in laser operation and
consider only the evolution of atoms exclusively coupled to their own
radiation field.

In solid state physics, the exciton-polariton state is\ one of the limiting
cases of superradiance. When a Frenkel exciton couples to the radiation
field in a small system which contains $N$ lattice points, it represents one
excited atom and $N-1$ unexcited atoms in the others. According to Dicke's
theory, the decay rate of the system will be enhanced by a factor $N$. But
as it was well known in a 3-D bulk crystal\cite{7},\textit{\ }the exciton
will couple with photon to form polariton--the eigenstate of the combined
system consisting of the crystal and the radiation field which does not
decay radiatively. What makes the exciton trapped in the bulk crystal is the
conservation of crystal momentum. If one considers a linear chain or a thin
film, the radiative recombination of excitons is fundamentally different
from the three-dimensional case. In bulk crystal, radiative decay requires
phonons or other translational-invariance breaking entities such as defects,
impurities, or interfaces. Thin films or linear chains, however, inherently
break translational invariance. As a consequence, a radiative decay channel
opens up, and the decay rate of the exciton is enhanced by a factor of $%
\lambda /d$ in a linear chain\cite{8} and ($\lambda /d)^{2}$ for 2D
exciton-polariton\cite{9,10}, where $\lambda $ is the wave length of emitted
photon and $d$ is the lattice constant of the linear chain or the thin film.

With the advances of microfabrication technologies such as molecular beam
epitaxy, it has become possible to fabricate various structures of
microcrystals with fine quality and novel properties, such as quantum well,
superlattice, quantum dot, quantum wire, and quantum ring. The exciton in a
quantum well can exhibit the behavior between purely three dimensions and
two dimensions. Many investigations on the radiative linewidth of excitons
in quantum wells have been performed \cite{11,12,13,14,15,16,17}. For lower
dimensional systems, first observation of superradiant short lifetimes of
excitons was performed by Ya. Aaviksoo \emph{et al}.\cite{18} on surface
states of the anthracene crystal. Superradiance has also been discussed for
one-dimensional (1D) Frenkel excitons in disordered aggregates.\cite{19,20}
A. L. Ivanov and H. Haug\cite{21} predicted the existence of exciton
crystal, which favors coherent emission in the form of superradiance, in
quantum wires. Y. Manabe \emph{et al}.\cite{22} considered the superradiance
of interacting Frenkel excitons in a linear chain. For quantum dots of CuCl
with radii R between 18 and 77 $\overset{\circ }{A}$, the superradiance of
excitons was also observed by Nakamura \emph{et al}.\cite{23}. The decay
rate was shown to be proportional to R$^{2.1}$ which confirms the
theoretical prediction by Hanamura.\cite{13} Similar works were also
obtained by Itch \emph{et al}.\cite{24}. By using numerical simulation,
Spano \emph{et al}. \cite{24} have also showed the effective coherent size
(which is inversely proportional to the decay time in their definition)
decreases as the temperature increases. Recently, superradiance of
polaritons from a dimer to finite one-dimensional crystal has also been
discussed by Dubovskii.\cite{26}

In reality, superradiance is accompanied by frequency shift, as pointed by
Lee \emph{et al}.\cite{27}. However, the coherent frequency shift of an
exciton has received fewer attention. One of the reasons is the difficulty
arose from the divergent nature of frequency shift, both infrared and
ultraviolet. Lee \emph{et al}. \cite{28} have solved the problem by using
the method of renormalization for a system of two atoms, and applied to the
case of excitons in a thin semiconductor film.\cite{27} Recently, we have
generalized their results to the quantum well systems.\cite{29} The
crossover behavior from 2D film to 3D crystal was also examined. It was
found that both the decay rate and renormalized frequency shift show
oscillatory dependence on layer thickness.

On the other hand, Purcell\cite{30} predicted that the spontaneous emission
rate, and thus the relaxation lifetime of an excited atom, would be altered
if the atom was put in a cavity with dimensions comparable to the transition
wavelength of the atom. The reason for modified spontaneous emission rate is
that spontaneous emission can be viewed as stimulated emission, stimulated
by vacuum-field fluctuations. Hence, the lifetime can be altered by
modifying the photon density of states.

In the last few years, experiments which have confined photons in
low-dimensional semiconductor quantum structures by using optical
microcavities were in progress. Researche on combined quantum confined 2D
carriers and 2D photon states has been performed by using quantum wells
embedded in planar microcavities, yielding interesting physics both in the
weak and strong carrier-photon coupling regimes.\cite{31} Results from 2D
carriers combined with 1D (0D) photon states have also been studied \cite%
{32,33}. Introducing further degrees of carrier confinement in such
microcavity structures is a natural trend in this field. Recently, the
incorporation of quantum dots in planar and pillar microcavities have been
reported.\cite{34} A coupling between 1D electron state and 2D photon states
were also obtained by inserting an array of quantum wires into a planar
microcavity.\cite{35} Such systems lead to dramatic changes in the
exciton-polariton dispersion\cite{36} and exciton lifetime.\cite{37,38}

In this manuscript, an alternative way is proposed to observe the inhibited
or enhanced spontaneous emission by embedding a quantum dot inside a
microcavity\cite{21}. After the injection of an electron and hole into a
quantum dot, a photon is generated by the recombination of the exciton. This
process allows one to determine Purcell effect by measuring the current
through the quantum dot. Similarly, by embedding two quantum dots inside the
cavity and controlling the gate voltage of one of the dots, one can not only
determine the superradiant effect by measuring the stationary current, but
also induces the entangled states, which is one of the fundamental
requirements for quantum information processing .

This paper is organized as follows. A brief description of superradiant and
Purcell effects is reviewed in section II. In section III, we discussed the
transport properties of a two level quantum dot embedded in a planar
microcavity. Electrical measurements of superradiance and the generated
entanglement are shown in section IV. Finally, overall conclusions are
presented in the last section.

\section{A brief review of Superradiant and Purcell effect}

\subsection{Spontaneous emission of two coupled atoms}

Spontaneous emission is one of the fundamental concepts in quantum mechanics
that can be traced back to the early works of Albert Einstein.\cite{39} In
free space a two level atom interacts with a continuum of radiation field
modes, which leads to an irreversible exponential decay of the excitation.
In this subsection, the Weisskopf-Wigner theory of spontaneous emission for
two level atoms in free space is reviewed.\cite{40}

The total Hamiltonian for a single two level atom in the rotating wave
approximation (RWA) is given by

\begin{equation}
H=\frac{1}{2}\hbar \omega \overset{\wedge }{\sigma }_{z}+\sum_{\mathbf{q}%
}\hbar \omega _{\mathbf{q}}(b_{\mathbf{q}}^{\dagger }b_{\mathbf{q}}+\frac{1}{%
2})-\sum_{\mathbf{q}}\hbar D_{\mathbf{q}}(b_{\mathbf{q}}^{\dagger }\overset{%
\wedge }{\sigma }_{-}+b_{\mathbf{q}}\overset{\wedge }{\sigma }_{+}),
\tag{2.1}
\end{equation}%
where $\overset{\wedge }{\sigma }_{z}=\left| \uparrow \right\rangle
\left\langle \uparrow \right| -\left| \downarrow \right\rangle \left\langle
\downarrow \right| $ and $\overset{\wedge }{\sigma }_{+}=\left| \uparrow
\right\rangle \left\langle \downarrow \right| ,$ $\overset{\wedge }{\sigma }%
_{-}=\left| \downarrow \right\rangle \left\langle \uparrow \right| $ are the
Pauli matrices in the $2\times 2$ space, $\hbar \omega $ is the level
spacing between the two levels, and $D_{\mathbf{q}}=(\frac{\omega _{\mathbf{q%
}}}{2\epsilon _{0}\hbar V})^{1/2}\overrightarrow{\epsilon }\cdot
\overrightarrow{\mu }$ is\ the dipole coupling matrix element. Furthermore, $%
\omega _{\mathbf{q}}=c\left| \mathbf{q}\right| ,$ and $b_{\mathbf{q}%
}^{\dagger }$ creates a photon with wave vector $\mathbf{q}$.

In the interaction picture, the combined atom-field system is represented by

\begin{equation}
\left| \psi (t)\right\rangle =f_{0}(t)\left| \uparrow ;0\right\rangle +\sum_{%
\mathbf{q}}f_{G;\mathbf{q}}(t)\left| \downarrow ;1_{q}\right\rangle ,
\tag{2.2}
\end{equation}%
where $\left| \downarrow ;1_{q}\right\rangle $ represents the state in which
the atom is in the ground state and the field mode $\mathbf{q}$ has one
photon.

Substituting Eq. (2.1) into Schr\"{o}dinger equation and projecting into
each state, we obtain

\begin{equation}
f_{0}(t)=\exp [-(\frac{1}{2}\Gamma +i\Delta \omega )t],  \tag{2.3}
\end{equation}%
where

\begin{eqnarray}
\Gamma &=&\frac{1}{3\pi \varepsilon _{0}\hbar c^{3}}\int d\omega _{\mathbf{q}%
}\omega _{\mathbf{q}}^{3}\mu ^{2}\delta (\omega -\omega _{\mathbf{q}})
\notag \\
&=&\frac{\omega ^{3}\mu ^{2}}{3\pi \varepsilon _{0}\hbar c^{3}},
\nonumber \hspace{0.745\textwidth} (2.4)
\end{eqnarray}%
and

\begin{equation}
\Delta \omega =\frac{1}{6\pi ^{2}}\varepsilon _{0}\hbar c^{3}\mathcal{P}\int
d\omega _{\mathbf{q}}\frac{\omega _{\mathbf{q}}^{3}\mu ^{2}}{\omega -\omega
_{\mathbf{q}}}.  \tag{2.5}
\end{equation}%
The spontaneous decay rate $\Gamma $ gives the Einstein's $A$ coefficient,
and the frequency shift $\Delta \omega $ represents the Lamb shift.\cite{41}

In a system of\emph{\ two identical atoms} interacting via common radiation
field, the decay splits into a sub- and a superradiant channel. The
Hamiltonian for two atoms interacting with the electromagnetic field reads

\begin{eqnarray}
H &=&H_{0}+H_{ph}+H_{eph}  \notag \\
H_{0} &=&\frac{1}{2}\hbar \omega (\overset{\wedge }{\sigma }_{1,z}+\overset{%
\wedge }{\sigma }_{2,z})  \notag \\
H_{ph} &=&\sum_{\mathbf{q}}\hbar \omega _{\mathbf{q}}(b_{\mathbf{q}%
}^{\dagger }b_{\mathbf{q}}+\frac{1}{2})  \notag \\
H_{eph} &=&\sum_{\substack{ \mathbf{q}  \\ j=1,2}}\hbar D_{\mathbf{q}}(b_{%
\mathbf{q}}^{\dagger }e^{i\mathbf{q\cdot r}_{j}}\overset{\wedge }{\sigma }%
_{j,-}+b_{\mathbf{q}}e^{-i\mathbf{q\cdot r}_{j}}\overset{\wedge }{\sigma }%
_{j,+}), \nonumber \hspace{0.35\textwidth} (2.6)
\end{eqnarray}%
where $\overset{\wedge }{\sigma }_{j,-},$ $\overset{\wedge }{\sigma }_{j,+}$%
and $\overset{\wedge }{\sigma }_{j,z}$ are the Pauli matrices in the $%
2\times 2$ space of the upper/lower level $\left| \uparrow \right\rangle
^{j},$ $\left| \downarrow \right\rangle ^{j}$ of atom $j$, $\hbar \omega $
is the level spacing between the two level, and $D_{\mathbf{q}}=(\frac{%
\omega _{\mathbf{q}}}{2\epsilon _{0}\hbar V})^{1/2}\overrightarrow{\epsilon }%
\cdot \overrightarrow{\mu }$ is\ the dipole coupling matrix element.
Furthermore, $\omega _{\mathbf{q}}=c\left| \mathbf{q}\right| ,$ and $b_{%
\mathbf{q}}^{\dagger }$ creates a photon with wave vector $\mathbf{q}$.

One can define the so-called Dicke state,

\begin{eqnarray}
\left| S_{0}\right\rangle &=&\frac{1}{\sqrt{2}}(\left| \uparrow \downarrow
\right\rangle -\left| \downarrow \uparrow \right\rangle )  \notag \\
\left| T_{1}\right\rangle &=&\left| \uparrow \uparrow \right\rangle  \notag
\\
\left| T_{0}\right\rangle &=&\frac{1}{\sqrt{2}}(\left| \uparrow \downarrow
\right\rangle +\left| \downarrow \uparrow \right\rangle )  \notag \\
\left| T_{-1}\right\rangle &=&\left| \downarrow \downarrow
\right\rangle . \nonumber \hspace{0.75\textwidth} (2.7)
\end{eqnarray}%
Using this basis, one can easily calculate the matrix elements

\begin{eqnarray}
\left\langle T_{1}\right| \overset{\wedge }{\sigma }_{j,\pm }\left|
T_{1}\right\rangle &=&\left\langle T_{1}\right| \overset{\wedge }{\sigma }%
_{j,\pm }\left| T_{-1}\right\rangle =0,\text{ }j=1,2  \notag \\
\left\langle T_{1}\right| \overset{\wedge }{\sigma }_{j,+}\left|
T_{0}\right\rangle &=&\left\langle T_{0}\right| \overset{\wedge }{\sigma }%
_{j,+}\left| T_{-1}\right\rangle =\frac{1}{\sqrt{2}},\text{ }j=1,2  \notag \\
\left\langle T_{0}\right| \overset{\wedge }{\sigma }_{j,\pm }\left|
S_{0}\right\rangle &=&0,\text{ }j=1,2  \notag \\
\left\langle T_{1}\right| \overset{\wedge }{\sigma }_{1,+}\left|
S_{0}\right\rangle &=&-\frac{1}{\sqrt{2}},\text{ }\left\langle T_{1}\right|
\overset{\wedge }{\sigma }_{2,+}\left| S_{0}\right\rangle =\frac{1}{\sqrt{2}}
\notag \\
\left\langle S_{0}\right| \overset{\wedge }{\sigma }_{1,+}\left|
T_{-1}\right\rangle &=&\frac{1}{\sqrt{2}},\text{ }\left\langle S_{0}\right|
\overset{\wedge }{\sigma }_{2,+}\left| T_{-1}\right\rangle =-\frac{1}{\sqrt{2%
}}.  \nonumber \hspace{0.35\textwidth} (2.8)
\end{eqnarray}

This means that there are two transition rate $\Gamma _{\pm }$ for
spontaneous of photons into a vacuum state,

\begin{equation}
\Gamma _{\pm }(q)=2\pi \sum_{\mathbf{q}}\frac{\left| \alpha _{\mathbf{q}}\pm
\beta _{\mathbf{q}}\right| ^{2}}{2}\delta (\omega -\omega _{q}),\text{ }q=%
\frac{\omega }{c},  \tag{2.9}
\end{equation}%
where we have defined $\alpha _{\mathbf{q}}=D_{\mathbf{q}}e^{i\mathbf{q\cdot
r}_{1}}$ and $\beta _{\mathbf{q}}=D_{\mathbf{q}}e^{i\mathbf{q\cdot r}_{2}}$.
Evaluation of this expression yields

\begin{equation}
\Gamma _{\pm }(q)=\Gamma \lbrack 1\pm \frac{\sin (qd)}{qd}],  \tag{2.10}
\end{equation}%
where $\Gamma \propto q^{3}$ is the decay rate of an isolated atom. Here, $%
d=\left| \mathbf{r}_{1}-\mathbf{r}_{2}\right| $ is the distance between the
two atoms. The appearance of two decay channels has been discovered by Dicke %
\cite{1} and observed by DeVoe and Brewer\cite{42} in a laser-trapped
two-ion system.

The time-dependence of the collective decay of two radiators is different
from the decay of two single radiators. If we denote the occupation
probabilities of the four levels by $T_{1}(t),$ $T_{0}(t),$ $T_{-1}(t),$ and
$S_{0}(t),$ the time dependence occupations is then governed by two decay
rates $\Gamma _{+}$ and $\Gamma _{-}$ :

\begin{eqnarray}
\overset{\cdot }{T_{1}} &=&-(\Gamma _{+}+\Gamma _{-})T_{1}  \notag \\
\overset{\cdot }{S_{0}} &=&\Gamma _{-}(T_{1}-S_{0})  \notag \\
\overset{\cdot }{T_{0}} &=&\Gamma _{+}(T_{1}-T_{0})  \notag \\
\overset{\cdot }{T_{-1}} &=&\Gamma _{+}T_{0}+\Gamma _{-}S_{0}.
\nonumber \hspace{0.64\textwidth} (2.11)
\end{eqnarray}%
The above equation can be solved easily,

\begin{eqnarray}
T_{1}(t) &=&e^{-(\Gamma _{+}+\Gamma _{-})t}  \notag \\
S_{0}(t) &=&\frac{[e^{-\Gamma _{-}t}-e^{-(\Gamma _{-}+\Gamma _{+})t}]\Gamma
_{-}}{\Gamma _{+}}  \notag \\
T_{0}(t) &=&\frac{[e^{-\Gamma _{+}t}-e^{-(\Gamma _{-}+\Gamma _{+})t}]\Gamma
_{+}}{\Gamma _{-}}  \nonumber \hspace{0.48\textwidth} (2.12) \\
T_{-1}(t) &=&\frac{\Gamma _{-}\Gamma _{+}-e^{-(\Gamma _{-}+\Gamma
_{+})t}[(-1+e^{\Gamma _{+}t})\Gamma _{-}^{2}+\Gamma _{-}\Gamma
_{+}+(-1+e^{\Gamma _{-}t})\Gamma _{+}^{2}]}{\Gamma _{-}\Gamma _{+}},
\notag
\end{eqnarray}%
where initial conditions $T_{1}(0)=1,$ $S_{0}(0)=T_{0}(0)=T_{-1}(0)=0$ have
been assumed. If we consider the special case where $\Gamma _{-}=0$ and $%
\Gamma _{+}=2\Gamma ,$ this would correspond to the case $qd\rightarrow 0$,
i.e. the wavelength of the emitted photon is much larger than the distance
between the two radiators. Then, Eq. (2.12) reduced to

\begin{eqnarray}
T_{1}(t) &=&e^{-\Gamma _{+}t}  \notag \\
T_{0}(t) &=&\Gamma _{+}te^{-\Gamma _{+}t}  \notag \\
S_{0}(t) &=&0  \notag \\
T_{-1}(t) &=&1-e^{-\Gamma _{+}t}(1+\Gamma _{+}t). \nonumber
\hspace{0.52\textwidth} (2.13)
\end{eqnarray}%
The total coherent emission rate $I_{2}(t)$ at time $t$ is the sum of the
emission rates from $T_{1}$ and $T_{0}$ :

\begin{equation}
I_{2}(t)=E_{0}\Gamma _{+}e^{-\Gamma _{+}t}(1+\Gamma _{+}t),\text{ }\Gamma
_{+}=2\Gamma ,  \tag{2.14}
\end{equation}%
where $E_{0}$ is a constant with dimension energy. This is different to the
incoherent sum $2I_{1}(t)$ of the emission rates $I_{1}(t)$ from two
independent radiators, which would give

\begin{equation}
2I_{1}(t)=2E_{0}\Gamma e^{-\Gamma t}.  \tag{2.15}
\end{equation}

\subsection{Effect of cavity on the radiative decay of excitons in low
dimensional systems}

As we mentioned above, the electron-hole pair is naturally a candidate for
examining spontaneous emission and Purcell effect. Let us first consider a
Wannier exciton in a quantum ring with radius $\rho \sim Nd/2\pi $, where $d$
is the lattice spacing and $N$ is the number of the lattice points. In our
model, the circular ring is joined by the $N$ lattice points, and we also
assume the effective mass approximation is valid in the circumference
direction. The state of the Wannier exciton can be specified as $\left| \nu
,n,m\right\rangle $, where $\nu $ is the exciton wave number. $n$ and $m$
are quantum numbers for internal structure of the exciton, and will be
specified later. Here, $\nu $ takes the value of an integer. The matter
Hamiltonian can be written as

\begin{equation}
H_{ex}=\sum_{\nu nm}E_{\nu nm}c_{\nu nm}^{\dagger }c_{\nu nm},  \tag{2.16}
\end{equation}%
where $c_{\nu nm}^{\dagger }$ and $c_{\nu nm}$ are the creation and
destruction operators of the exciton, respectively. The Hamiltonian of free
photon is

\begin{equation}
H_{ph}=\sum_{\mathbf{q}^{\prime }k_{z}^{\prime }}\hbar c(q^{\prime
2}+k_{z}^{\prime 2})^{1/2}b_{\mathbf{q}^{\prime }k_{z}^{\prime }\lambda
}^{\dagger }b_{\mathbf{q}^{\prime }k_{z}^{\prime }\lambda },  \tag{2.17}
\end{equation}%
where $b_{\mathbf{q}^{\prime }k_{z}^{\prime }\lambda }^{\dagger }$ and $b_{%
\mathbf{q}^{\prime }k_{z}^{\prime }\lambda }$ are the creation and
destruction operators of the photon, respectively. The wave vector $\mathbf{k%
}^{\prime }$\hspace{0.06in}of the photon is separated into two parts: $%
k_{z}^{\prime }$ is the perpendicular component of $\mathbf{k}^{\prime }$ on
the ring plane such that $k^{\prime 2}=q^{\prime 2}+k_{z}^{\prime 2}$.

The interaction between the exciton and the photon can be expressed as

\begin{equation}
H^{\prime }=\sum_{k_{z}^{\prime }nm}\sum_{\mathbf{q}^{\prime }}D_{\mathbf{q}%
^{\prime }k_{z}^{\prime }\nu nm}b_{k_{z}^{\prime }\mathbf{q}^{\prime
}}c_{\nu nm}^{\dagger }+\mathbf{h.c.,}  \tag{2.18}
\end{equation}%
where

\begin{equation}
D_{\mathbf{q}^{\prime }k_{z}^{\prime }\nu nm}=H_{\nu }^{(1)}(q^{\prime }\rho
)\frac{e}{mc}\sqrt{\frac{2\pi \hbar c}{(q^{\prime 2}+k_{z}^{\prime 2})^{1/2}v%
}}\epsilon _{\mathbf{q}^{\prime }k_{z}^{\prime }}A_{\nu nm}  \tag{2.19}
\end{equation}%
with $\mathbf{\epsilon }_{\mathbf{q}^{\prime }k_{z}^{\prime }}$ being the
polarization of the photon and $H_{\nu }^{(1)}$ is the Hankel function. In
Eq. (2.19),

\begin{eqnarray}
A_{\nu nm} &=&\sqrt{N}\sum_{\varphi _{e}}F_{nm}(\varphi _{e})\int d\mathbf{%
\varphi }w_{c}(\varphi -\varphi _{e})  \notag \\
&&\times \exp (i\nu (\varphi -\frac{m_{e}^{\ast }\varphi
_{e}}{m_{e}^{\ast }+m_{h}^{\ast }}))(-i\hbar \frac{\partial
}{\partial \varphi })w_{v}(\varphi ). \nonumber
\hspace{0.30\textwidth} (2.20)
\end{eqnarray}%
is the effective transition dipole matrix element and $F_{nm}(\varphi _{e})$
is the hydrogenic wavefunction in the ring. Here, $m_{e}^{\ast }$ and $%
m_{h}^{\ast }$ \ are, respectively, the effective masses of the electron and
hole.

The decay rate of the exciton can be expressed as

\begin{equation}
\gamma _{_{\nu nm}}=2\pi \sum_{\mathbf{q}^{\prime }k_{z}^{\prime }\lambda
}\left| D_{\mathbf{q}^{\prime }k_{z}^{\prime }\nu nm}\right| ^{2}\delta
(\omega _{\mathbf{q}^{\prime }k_{z}^{\prime }\nu nm}),  \tag{2.21}
\end{equation}%
where $\omega _{\mathbf{q}^{\prime }k_{z}^{\prime }\nu nm}=E_{\nu nm}/\hbar
-c\sqrt{q^{\prime 2}+k_{z}^{\prime 2}}.$ The Wannier exciton decay rate in
the optical region can be calculated straightforwardly and is given by

\begin{equation}
\gamma _{\nu nm}=\frac{e^{2}\hbar }{m^{2}c}\frac{\rho }{d}\int \left| H_{\nu
}^{(1)}(q^{\prime }\rho )\right| ^{2}q^{\prime }\int \frac{\delta (\omega _{%
\mathbf{q}^{\prime }k_{z}^{\prime }\nu nm})}{\sqrt{k_{z}^{\prime
2}+q^{\prime 2}}}\left| \mathbf{\epsilon }_{\mathbf{q}^{\prime
}k_{z}^{\prime }\lambda }\cdot \mathbf{\chi }_{\nu nm}\right|
^{2}dk_{z}^{\prime }dq^{\prime },  \tag{2.22}
\end{equation}%
where

\begin{equation}
\mathbf{\chi }_{\nu nm}=\sum_{\varphi _{e}}F_{nm}^{\ast }(\varphi _{e})\int d%
\mathbf{\varphi }w_{c}^{\ast }(\varphi -\varphi _{e})(-i\hbar \frac{\partial
}{\partial \varphi })w_{v}(\mathbf{\varphi }).  \tag{2.23}
\end{equation}%
\begin{figure}[th]
\center\includegraphics[width=8cm]{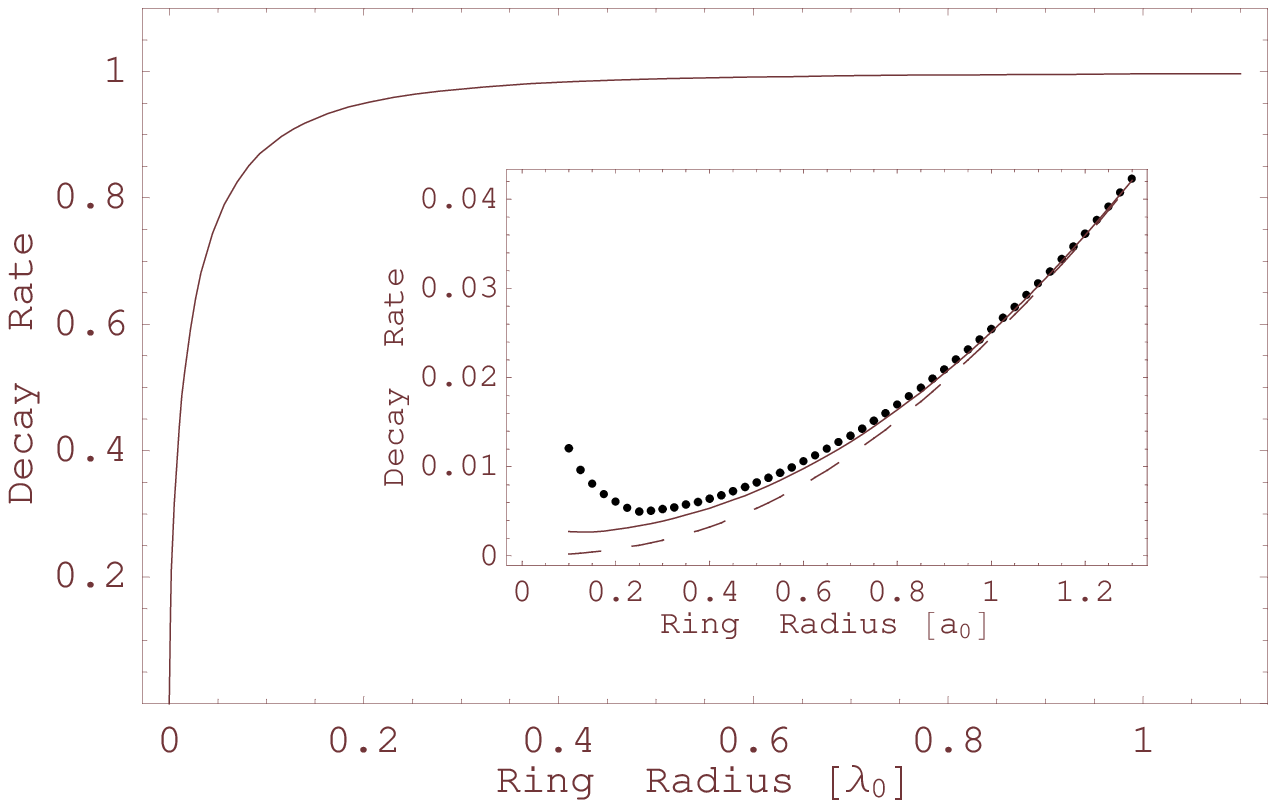}
\caption{Decay rate of a\ quantum ring exciton in large radius limit, i.e. $%
F_{nm}^{\ast }$ are assumed to be independent of $\protect\rho $. The
vertical unit and horizontal units are $\frac{3\protect\pi }{2k_{0}d}\protect%
\gamma _{0}$ and $\protect\lambda _{0}$, respectively. Inset : Effect of
Aharonov-Bohm on the radiative decay of quantum ring exciton. The dashed(--
--), solid, and dashed-dotted(-- $\cdot $) curves correspond to $\Phi =0\Phi
_{0},$ $0.25\Phi _{0},$ and $0.5\Phi _{0},$ respectively. In small radius
limit, $F_{nm}^{\ast }$ depends strongly on radius $\protect\rho $, and its
influence on the decay rate is evident. The vertical and horizontal units
here are $\frac{3\protect\pi }{2k_{0}d}\protect\gamma _{0}$ and ring radius
(in units of $a_{0}$), respectively. }
\end{figure}

From Eq. (2.22), one observes that the decay rate $\gamma _{\nu nm}$ is
proportional to $\rho /d$. This is just the superradiance factor coming from
the coherent contributions of atoms within half a wavelength or so. In Fig.
1 we have numerically calculated the superradiant decay rate in $\nu =0,n=0$%
, and $m=0$ mode. In plotting the figure, we have assumed $E_{\nu nm}/\hbar
=k_{0}=2\pi /\lambda ,$ $\ \lambda =8000\overset{\circ }{A}$, $d=5\overset{%
\circ }{A},\gamma _{0}$ is the decay rate of an isolated atom, and for large
radius, $F_{nm}^{\ast }$ is independent of $\rho .$ The decay rate increases
linearly with the increasing of radius when the radius is small. This linear
regime agrees with Dicke's prediction: For one excited atom and $N-1$
unexcited atoms in a small volume, the decay rate is enhanced by the factor
of $1\times N$. For large radius, the decay rate can approach 1D limit $(=%
\frac{3\pi }{2k_{0}d}\gamma _{0})$ correctly.
\begin{figure}[th]
\center\includegraphics[width=8cm]{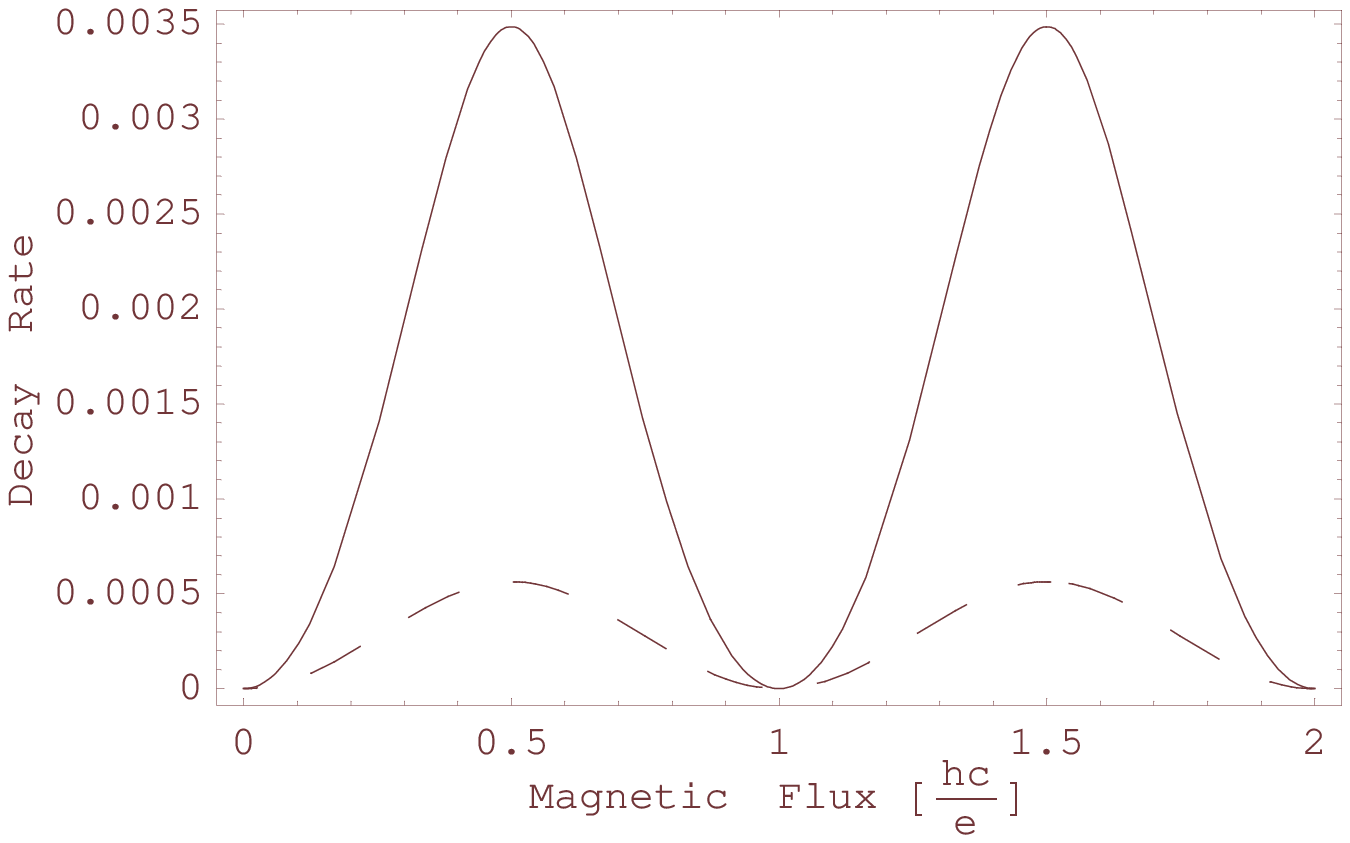}
\caption{Dependence of relative decay rate [$\protect\gamma _{_{\protect\nu %
nm}}(\Phi )-\protect\gamma _{_{\protect\nu nm}}(\Phi =0)$] on the magnetic
flux. The dashed and solid curves correspond to $\protect\rho =0.5$ $a_{0}$
and $\protect\rho =1$ $a_{0},$ respectively. The vertical and horizontal
units are $\frac{3\protect\pi }{2k_{0}d}\protect\gamma _{0}$ and universal
flux quantum $\Phi _{0}=hc/e,$ respectively.}
\end{figure}

Quite recently, R. A. R\"{o}mer and M. E. Raikh studied theoretically the
exciton absorption shredded by a magnetic flux $\Phi $\cite{43}. From their
results, effects of magnetic flux on exciton wavefunction $F_{nm}^{\ast }$
can not be neglected in small radius limit, and may be examined from the
variations of the decay rate. In the inset of Fig. 1, three curves of
different flux $\Phi $ are presented as functions of radius $\rho $. The
dashed, solid, and dotted curves represent the cases of $\Phi =0\Phi _{0},$ $%
0.25\Phi _{0},$ and $0.5\Phi _{0},$ respectively. For $\Phi =0.5\Phi _{0}$,
the decay rate decreases as the ring radius becomes small but reaches the
minimum point as $\rho $ is about 0.25$a_{0}$(where $a_{0}$ is the effective
Bohr radius we assumed in 1D limit). This is because the probability, for
electron and hole to meet each other on the opposite side of the ring,
increases with the decreasing of ring radius, while the coherent effect
decreases with the decreasing of the radius. Therefore, there is a
competition between these two effects as one decreases the radius. In Fig.
2, relative decay rate [$\gamma _{_{\nu nm}}(\Phi )-\gamma _{_{\nu nm}}(\Phi
=0)$] is plotted as a function of magnetic flux $\Phi $ with different
radius. The solid and dashed lines represent the cases of $\rho =1$ $a_{0}$
and $\rho =0.5$ $a_{0}$ , respectively. As expected, the larger the radius,
the smaller the AB oscillation amplitude. Besides, the superradiant decay
rate is most enhanced for $\Phi =0.5\Phi _{0}$, and the oscillation period
is equal to $\Phi _{0}=hc/e$.

We now consider a Wannier exciton in a quantum ring embedded in perfectly
reflecting mirrors with cavity length $L_{c}$. The decay rate of the quantum
ring exciton can be expressed as

\begin{equation}
\gamma _{\nu }=\sum_{n_{c}}\frac{e^{2}\hbar }{m^{2}c^{2}L_{c}}\frac{\rho }{d}%
\left| H_{\nu }^{(1)}(\sqrt{(2\pi /\lambda )^{2}-(\pi n_{c}/L_{c})^{2}}\rho
)\right| ^{2}\left| \mathbf{\epsilon }_{\mathbf{q}^{\prime }k_{z}^{\prime
}}\cdot \mathbf{\chi }_{\nu }\right| ^{2}.  \tag{2.24}
\end{equation}%
The numerical calculations of Eq. (2.24) are shown in the left panel of Fig.
3. As can be seen, the decay rate of a quantum ring exciton shows the
enhanced peaks as the cavity length $L_{c}$ is equal to multiple
half-wavelengths of the emitted photon.
\begin{figure}[th]
\center\includegraphics[width=8cm]{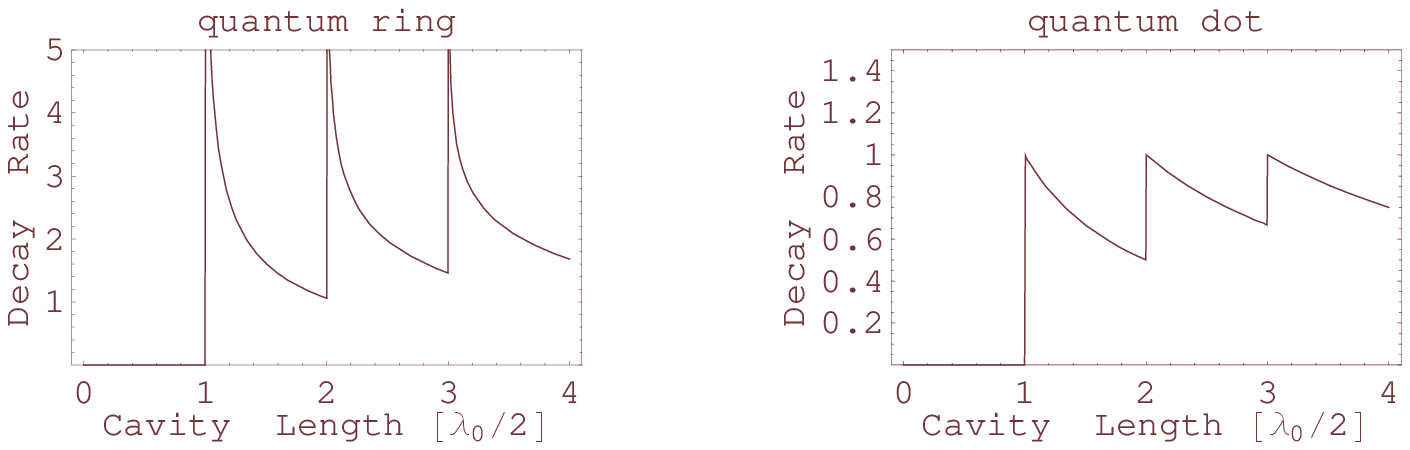} \caption{Left panel :
Decay rate of a quantum ring exciton in a planar microcavity with
radius $\protect\rho =\protect\lambda /2\protect\pi $. The
horizontal and vertical units are ($\protect\lambda /2$) and $\frac{%
e^{2}\hbar \protect\lambda ^{2}}{4\protect\pi ^{3}m^{2}c^{2}d}\left| \mathbf{%
\protect\epsilon }_{\mathbf{q}^{\prime }k_{z}^{\prime }}\cdot \mathbf{%
\protect\chi }_{\protect\nu }\right| ^{2},$ respectively. Right Panel :
Similar case for a quantum dot exciton.}
\end{figure}

However, if one considers a quantum dot exciton inside the microcavity, the
decay rate reads

\begin{equation}
\gamma \propto \sum_{n_{c}}\frac{e^{2}\hbar }{m^{2}c^{2}L_{c}}\theta ((2\pi
/\lambda )^{2}-(\pi n_{c}/L_{c})^{2})\left| \mathbf{\epsilon }_{\mathbf{q}%
^{\prime }k_{z}^{\prime }}\cdot \mathbf{\chi }\right| ^{2},  \tag{2.25}
\end{equation}%
where $\theta $ is the step function. The numerical calculations are
presented in the right panel of Fig. 3. One can see from the figure, there
is no enhanced peak with the increasing of the cavity length. This is
because the angular momentum (translational momentum) of the exciton in a
quantum ring is conserved in circular direction, while the crystal symmetry
is totally broken in a quantum dot. Due to the modification of the density
of states of the photon in the microcavity, the decay rate of the exciton
shows enhanced peaks in 1D systems\cite{38} and zigzag structure in 0D
quantum dot. One also notes that such kind of peak maybe a useful feature to
realize the Aharonov-Bohm effect for an exciton in a quantum ring. Generally
speaking, the excitonic AB oscillation is very small and hard to be
measured. However, if one can incorporate the quantum ring inside the planar
microcavities, the AB oscillation may be enhanced at these peaks.

\section{Current through one quantum dot and Purcell effect}

We now consider a quantum dot embedded in a \textit{p-i-n} junction which is
similar to the device proposed by O. Benson \textit{et al}\cite{44}. The
energy-band diagram is shown in Fig. 4.

\begin{figure}[th]
\center\includegraphics[width=8cm]{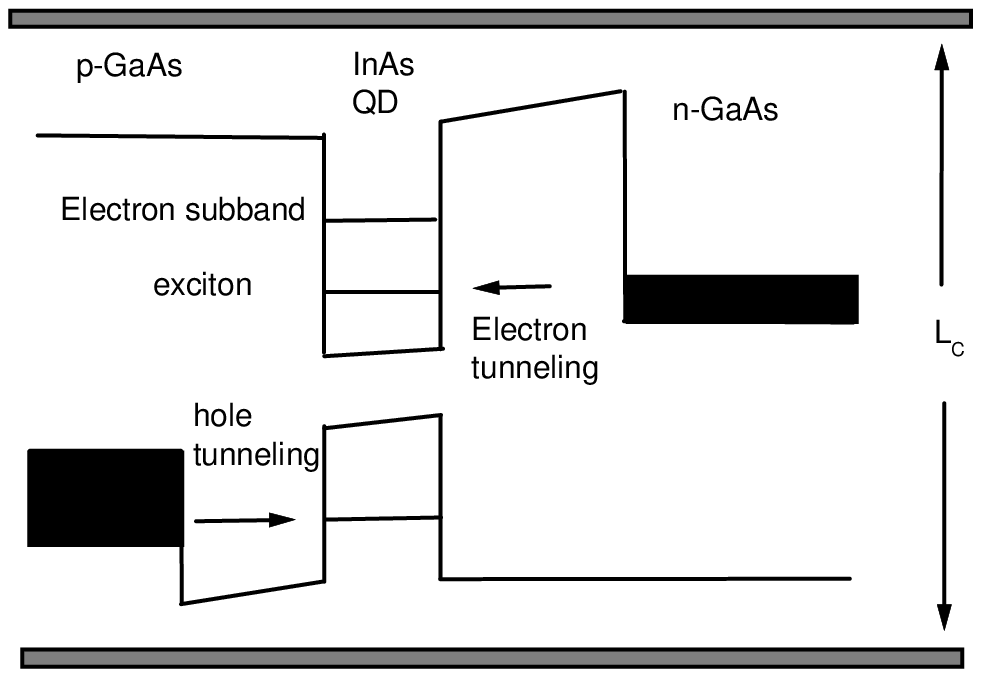} \caption{Energy-band
diagram of the p-i-n junction.}
\end{figure}

Both the hole and electron reservoirs are assumed to be in thermal
equilibrium. For the physical phenomena we are interested in, the fermi
level of the \textit{p(n)}-side hole (electron) is slightly lower (higher)
than the hole (electron) subband in the dot. After a hole is injected into
the hole subband in the quantum dot, the \textit{n}-side electron can tunnel
into the exciton level because of the Coulomb interaction between the
electron and hole. Thus, we may assume three dot states

\begin{eqnarray}
\left| 0\right\rangle &=&\left| 0,h\right\rangle  \notag \\
\left| U\right\rangle &=&\left| e,h\right\rangle  \notag \\
\text{ }\left| D\right\rangle &=&\left| 0,0\right\rangle \nonumber
\hspace{0.75\textwidth} (3.1)
\end{eqnarray}%
, where $\left| 0,h\right\rangle $ means there is one hole in the quantum
dot,\ $\left| e,h\right\rangle $ is the exciton state, and $\left|
0,0\right\rangle $ represents the ground state with no hole and electron in
the quantum dot. One might argue that one can not neglect the state $\left|
e,0\right\rangle $ for real device since the tunable variable is the applied
voltage. This can be resolved by fabricating a thicker barrier on the
electron side so that the probability for an electron to tunnel in advance
is very small. Moreover, the charged exciton and biexcitons states are also
neglected in our calculations. This means a low injection limit is required
in the experiment\cite{45}. We can now define the dot-operators $\overset{%
\wedge }{n_{U}}\equiv \left| U\right\rangle \left\langle U\right| ,$ $%
\overset{\wedge }{n_{D}}\equiv \left| D\right\rangle \left\langle D\right| ,$
$\overset{\wedge }{p}\equiv \left| U\right\rangle \left\langle D\right| ,$ $%
\overset{\wedge }{s_{U}}\equiv \left| 0\right\rangle \left\langle U\right| ,$
$\overset{\wedge }{s_{D}}\equiv \left| 0\right\rangle \left\langle D\right| $%
. The total hamiltonian $H$ of the system consists of three parts: the dot
hamiltonian, the photon bath, and the electron (hole) reservoirs:

\begin{eqnarray}
H &=&H_{0}+H_{T}+H_{V}  \notag \\
H_{0} &=&\varepsilon _{U}\overset{\wedge }{n_{U}}+\varepsilon _{D}\overset{%
\wedge }{n_{D}}+H_{p}+H_{res}  \notag \\
H_{T} &=&\sum_{k}g(D_{k}b_{k}^{\dagger }\overset{\wedge }{p}+D_{k}^{\ast
}b_{k}\overset{\wedge }{p}^{\dagger })=g(\overset{\wedge }{p}X+\overset{%
\wedge }{p}^{\dagger }X^{\dagger })  \notag \\
H_{p} &=&\sum_{k}\omega _{k}b_{k}^{\dagger }b_{k}  \notag \\
H_{V} &=&\sum_{\mathbf{q}}(V_{\mathbf{q}}c_{\mathbf{q}}^{\dagger }\overset{%
\wedge }{s_{U}}+W_{\mathbf{q}}d_{\mathbf{q}}^{\dagger }\overset{\wedge }{%
s_{D}}+c.c.)  \notag \\
H_{res} &=&\sum_{\mathbf{q}}\varepsilon _{\mathbf{q}}^{U}c_{\mathbf{q}%
}^{\dagger }c_{\mathbf{q}}+\sum_{\mathbf{q}}\varepsilon _{\mathbf{q}}^{D}d_{%
\mathbf{q}}^{\dagger }d_{\mathbf{q}}. \nonumber
\hspace{0.50\textwidth} (3.2)
\end{eqnarray}%
In above equations, $b_{k}$ is the photon operator, $gD_{k}$ is the dipole
coupling strength, $X=\sum_{k}D_{k}b_{k}^{\dagger }$ ,$\ $and $c_{\mathbf{q}%
} $ and $d_{\mathbf{q}}$ denote the electron operators in the left ad right
reservoirs, respectively. Here, $g$ is a constant with a unit of the
tunneling rate. The couplings to the electron and hole reservoirs are given
by the standard tunnel hamiltonian $H_{V},$ where $V_{\mathbf{q}}$ and $W_{%
\mathbf{q}}$ couple the channels $\mathbf{q}$ of the electron and the hole
reservoirs. If the couplings to the electron and the hole reservoirs are
weak, then it is reasonable to assume that the standard Born-Markov
approximation with respect to these couplings is valid. In this case, we
will derive a master equation from the exact time-evolution of the system.

In interaction picture, time evolutions of arbitrary operators $\overset{%
\wedge }{O}$ and $X$ are defined by

\begin{equation}
\widetilde{O}(t)\equiv e^{iH_{0}t}Oe^{-iH_{0}t},\text{ }X_{t}\equiv
e^{iH_{0}t}Xe^{-iH_{0}t}.  \tag{3.3}
\end{equation}%
Furthermore, for the total density matrix $\Xi (t)$ which obeys the
Liouville equation

\begin{equation}
\Xi (t)=e^{-iHt}\Xi _{t=0}e^{iHt},  \tag{3.4}
\end{equation}%
and we also define

\begin{equation}
\widetilde{\Xi }(t)\equiv e^{iH_{0}t}\Xi (t)e^{-iH_{0}t}.  \tag{3.5}
\end{equation}%
The expectation value of any operator $\overset{\wedge }{O}$ is given by

\begin{equation}
\overset{\wedge }{\left\langle O\right\rangle }_{t}\equiv Tr(\Xi (t)O)=Tr(%
\widetilde{\Xi }(t)\widetilde{O}(t)).  \tag{3.6}
\end{equation}%
We therefore have

\begin{eqnarray}
\widetilde{n_{U}}(t) &=&\overset{\wedge }{n_{U}},\text{ }\widetilde{n_{D}}%
(t)=\overset{\wedge }{n_{D}}  \notag \\
\widetilde{p}(t) &=&\overset{\wedge }{p}e^{i\varepsilon t}X_{t},\text{ }%
\widetilde{p}^{\dagger }(t)=\overset{\wedge }{p}^{\dagger }e^{-i\varepsilon
t}X_{t}^{\dagger }  \notag \\
\varepsilon &\equiv &\varepsilon _{U}-\varepsilon _{D}. \nonumber
\hspace{0.70\textwidth} (3.7)
\end{eqnarray}%
The equation of motion for $\widetilde{\Xi }(t)$ becomes

\begin{equation}
i\frac{d}{dt}\widetilde{\Xi }(t)=[\widetilde{H_{T}}(t)+\widetilde{H_{V}}(t),%
\widetilde{\Xi }(t)].  \tag{3.8}
\end{equation}%
This can be written as

\begin{eqnarray}
\frac{d}{dt}\widetilde{\Xi }(t) &=&-i[\widetilde{H_{T}}(t),\widetilde{\Xi }%
(t)]-i[\widetilde{H_{V}}(t),\widetilde{\Xi }(t)]  \notag \\
&=&-i[\widetilde{H_{T}}(t),\widetilde{\Xi }(t)]-i[\widetilde{H_{V}}(t),\Xi
_{0}]  \notag \\
&&-\int_{0}^{t}dt^{\prime }[\widetilde{H_{V}}(t),[\widetilde{H_{T}}%
(t^{\prime })+\widetilde{H_{V}}(t^{\prime }),\widetilde{\Xi
}(t^{\prime })]]. \nonumber \hspace{0.3\textwidth} (3.9)
\end{eqnarray}%
Now, we define the effective density operator of the dot plus photons,

\begin{equation}
\widetilde{\rho }(t)=Tr_{res}\widetilde{\Xi }(t)  \tag{3.10}
\end{equation}%
as the trace of $\widetilde{\Xi }(t)$ over electron reservoirs. The trace Tr$%
_{res}$ over the terms linear which are in $H_{V}$ vanishes, therefore,

\begin{equation}
\frac{d}{dt}\widetilde{\rho }(t)=-i[\widetilde{H_{T}}(t),\widetilde{\rho }%
(t)]-Tr_{res}\int_{0}^{t}dt^{\prime }[\widetilde{H_{V}}(t),[\widetilde{H_{V}}%
(t^{\prime }),\widetilde{\Xi }(t^{\prime })]].  \tag{3.11}
\end{equation}%
As can be seen from the above equation, the last term is already second
order in $H_{V},$ we can approximate

\begin{equation}
\widetilde{\Xi }(t^{\prime })\approx R_{0}\widetilde{\rho }(t^{\prime }),
\tag{3.12}
\end{equation}%
where $R_{0}$ is the equilibrium density matrix for the two electron
reservoirs. Working out the commutators and using the time evolution of the
electron reservoir operators

\begin{equation}
\widetilde{c_{\mathbf{q}}}(t)=e^{-i\varepsilon _{\mathbf{q}}^{L}t}c_{q},%
\text{ }\widetilde{d_{\mathbf{q}}}(t)=e^{-i\varepsilon _{\mathbf{q}%
}^{R}t}d_{q},  \tag{3.13}
\end{equation}%
the master equation becomes

\begin{eqnarray}
\widetilde{\rho }(t) &=&\rho _{0}-i\int_{0}^{t}dt^{\prime }[\widetilde{H_{T}}%
(t^{\prime }),\widetilde{\rho }(t^{\prime })]  \notag \\
&&-\Gamma _{L}\int_{0}^{t}dt^{\prime }\{\widetilde{s_{U}}(t^{\prime })%
\widetilde{s_{U}}^{\dagger }(t^{\prime })\widetilde{\rho }(t^{\prime })-2%
\widetilde{s_{U}}^{\dagger }(t^{\prime })\widetilde{\rho }(t^{\prime })%
\widetilde{s_{U}}(t^{\prime })\}  \notag \\
&&-\Gamma _{L}\int_{0}^{t}dt^{\prime }\{\widetilde{\rho }(t^{\prime })%
\widetilde{s_{U}}(t^{\prime })\widetilde{s_{U}}^{\dagger }(t^{\prime })\}
\notag \\
&&-\Gamma _{R}\int_{0}^{t}dt^{\prime }\{\widetilde{s_{D}}^{\dagger
}(t^{\prime })\widetilde{s_{D}}(t^{\prime })\widetilde{\rho }(t^{\prime })\}
\notag \\
&&-\Gamma _{R}\int_{0}^{t}dt^{\prime }\{-2\widetilde{s_{D}}(t^{\prime })%
\widetilde{\rho }(t^{\prime })\widetilde{s_{D}}^{\dagger }(t^{\prime })+%
\widetilde{\rho }(t^{\prime })\widetilde{s_{D}}^{\dagger }(t^{\prime })%
\widetilde{s_{D}}(t^{\prime })\}, \nonumber \hspace{0.15\textwidth}
(3.14)
\end{eqnarray}%
where $\Gamma _{L}=\pi \sum_{\mathbf{q}}V_{\mathbf{q}}^{2}\delta
(\varepsilon _{U}-\varepsilon _{\mathbf{q}}^{L})$ and $\Gamma _{R}=\pi \sum_{%
\mathbf{q}}W_{\mathbf{q}}^{2}\delta (\varepsilon _{D}-\varepsilon _{\mathbf{q%
}}^{R}).$

Multiplying Eq. (3.14) by $\overset{\wedge }{n_{U}},$ $\overset{\wedge }{%
n_{D}},$ $\overset{\wedge }{p},$ and $\overset{\wedge }{p}^{\dagger },$
respectively and performing the trace with the three dot states in Eq.
(3.1), one obtains

\begin{equation*}
\overset{\wedge }{\left\langle n_{U}\right\rangle }_{t}-\overset{\wedge }{%
\left\langle n_{U}\right\rangle }_{0}=-ig\int_{0}^{t}dt^{\prime }\{\overset{%
\wedge }{\left\langle p\right\rangle }_{t^{\prime }}-\overset{\wedge }{%
\left\langle p^{\dagger }\right\rangle }_{t^{\prime }}\}+2\Gamma
_{U}\int_{0}^{t}dt^{\prime }(1-\overset{\wedge }{\left\langle
n_{U}\right\rangle }_{t^{\prime }}-\overset{\wedge }{\left\langle
n_{D}\right\rangle }_{t^{\prime }})
\end{equation*}

\begin{equation*}
\overset{\wedge }{\left\langle n_{D}\right\rangle }_{t}-\overset{\wedge }{%
\left\langle n_{D}\right\rangle }_{0}=-ig\int_{0}^{t}dt^{\prime }\{\overset{%
\wedge }{\left\langle p\right\rangle }_{t^{\prime }}-\overset{\wedge }{%
\left\langle p^{\dagger }\right\rangle }_{t^{\prime }}\}-2\Gamma
_{D}\int_{0}^{t}dt^{\prime }\overset{\wedge }{\left\langle
n_{D}\right\rangle }_{t^{\prime }}
\end{equation*}

\begin{eqnarray*}
\overset{\wedge }{\left\langle p\right\rangle }_{t}-\overset{\wedge }{%
\left\langle p\right\rangle _{t}^{0}} &=&-\Gamma _{D}\int_{0}^{t}dt^{\prime
}e^{i\varepsilon (t-t^{\prime })}\left\langle X_{t}X_{t^{\prime }}^{\dagger }%
\widetilde{p}(t^{\prime })\right\rangle _{t^{\prime }} \\
&&-ig\int_{0}^{t}dt^{\prime }e^{i\varepsilon (t-t^{\prime })}\{\left\langle
\overset{\wedge }{n_{U}}X_{t}X_{t^{\prime }}^{\dagger }\right\rangle
_{t^{\prime }}-\left\langle \overset{\wedge }{n_{D}}X_{t^{\prime }}^{\dagger
}X_{t}\right\rangle _{t^{\prime }}\}
\end{eqnarray*}

\begin{eqnarray}
\overset{\wedge }{\left\langle p^{\dagger }\right\rangle }_{t}-\overset{%
\wedge }{\left\langle p\right\rangle _{t}^{0}} &=&-\Gamma
_{D}\int_{0}^{t}dt^{\prime }e^{-i\varepsilon (t-t^{\prime })}\left\langle
\widetilde{p}^{\dagger }(t^{\prime })X_{t^{\prime }}X_{t}^{\dagger
}\right\rangle _{t^{\prime }}  \notag \\
&&+ig\int_{0}^{t}dt^{\prime }e^{-i\varepsilon (t-t^{\prime
})}\{\left\langle \overset{\wedge }{n_{U}}X_{t^{\prime
}}X_{t}^{\dagger }\right\rangle _{t^{\prime }}-\left\langle
\overset{\wedge }{n_{D}}X_{t}^{\dagger }X_{t^{\prime }}\right\rangle
_{t^{\prime }}\}, \nonumber \hspace{0.08\textwidth} (3.15)
\end{eqnarray}%
where $\varepsilon =\varepsilon _{U}-\varepsilon _{D}$ is the energy gap of
the quantum dot exciton. Here, $\widetilde{p}(t^{\prime })=pe^{i\varepsilon
t}X_{t^{\prime }}$, and $X_{t^{\prime }}$ denotes the time evolution of $X$
with $H_{p}$. The expectation value $\overset{\wedge }{\left\langle
p^{(\dagger )}\right\rangle _{t}^{0}}$describes the decay of an initial
polarization of the system and plays no role for the stationary current.
Therefore, we shall assume the initial expectation value of $\overset{\wedge
}{p}^{(\dagger )}$ vanishes at time $t=0$.

As can be seen from Eqs. (3.15), there are terms like $\left\langle \overset{%
\wedge }{n_{U}}X_{t}X_{t^{\prime }}^{\dagger }\right\rangle _{t^{\prime }}$
which contain products of dot operators and photon operators. If we are
interested in small coupling parameters here, a decoupling of the reduced
density matrix $\widetilde{\rho }(t^{\prime })$ can be written as

\begin{equation}
\widetilde{\rho }(t^{\prime })\approx \rho _{ph}^{0}Tr_{ph}\widetilde{\rho }%
(t^{\prime }).  \tag{3.16}
\end{equation}%
By using above equation, we obtain

\begin{equation}
Tr(\widetilde{\rho }(t^{\prime })\overset{\wedge }{n_{U}}X_{t}X_{t^{\prime
}}^{\dagger })\approx \overset{\wedge }{\left\langle n_{U}\right\rangle }%
_{t^{\prime }}\left\langle X_{t}X_{t^{\prime }}^{\dagger }\right\rangle _{0}
\tag{3.17}
\end{equation}%
and correspondingly the other products of operators can be obtained also.
For spontaneous emission, the photon bath is assumed to be in equilibrium.
The expectation value $\left\langle X_{t}X_{t^{\prime }}^{\dagger
}\right\rangle _{0}\equiv C(t-t^{\prime })$ is a function of the time
interval only. We can now define the Laplace transformation for real $z,$

\begin{eqnarray}
\overset{}{C_{\varepsilon }}(z) &\equiv &\int_{0}^{\infty
}dte^{-zt}e^{i\varepsilon t}C(t)  \notag \\
\overset{}{n_{U}}(z) &\equiv &\int_{0}^{\infty }dte^{-zt}\overset{\wedge }{%
\left\langle n_{U}\right\rangle }_{t}\text{ \ }etc.,\text{ }z>0
\nonumber \hspace{0.44\textwidth} (3.18)
\end{eqnarray}%
and transform the whole equations of motion into $z$-space,

\begin{eqnarray}
n_{U}(z) &=&-i\frac{g}{z}(p(z)-p^{\ast }(z))+2\frac{\Gamma _{U}}{z}%
(1/z-n_{U}(z)-n_{D}(z))  \notag \\
n_{D}(z) &=&\frac{g}{z}(p(z)-p^{\ast }(z))-2\frac{\Gamma _{D}}{z}n_{D}(z)
\notag \\
p(z) &=&-ig\{n_{U}(z)C_{\varepsilon }(z)-n_{D}(z)C_{-\varepsilon }^{\ast
}(z)\}-\Gamma _{D}p(z)C_{\varepsilon }(z)  \notag \\
p^{\ast }(z) &=&ig\{n_{U}(z)C_{\varepsilon }^{\ast
}(z)-n_{D}(z)C_{-\varepsilon }(z)\}-\Gamma _{D}p^{\ast
}(z)C_{\varepsilon }^{\ast }(z). \nonumber \hspace{0.2\textwidth}
(3.19)
\end{eqnarray}%
These equations can then be solved algebraically. The tunnel current $%
\widehat{I}$ can be defined as the change of the occupation of $\overset{%
\wedge }{n_{U}}$ and is given by $\widehat{I}\equiv ig(\overset{\wedge }{p}-%
\overset{\wedge }{p}^{\dagger }),$ where we have set the electron charge $%
e=1 $ for convenience. The time dependence of the expectation value $\overset%
{\wedge }{\left\langle I\right\rangle }_{t}$ can be obtained by solving Eqs.
(3.19) and performing the inverse Laplace transformation. For time $%
t\rightarrow \infty ,$ the result is

\begin{eqnarray}
\overset{\wedge }{\left\langle I\right\rangle }_{t\rightarrow \infty } &=&%
\frac{2g^{2}\Gamma _{U}\Gamma _{D}B}{g^{2}\Gamma _{D}B+[g^{2}B+\Gamma
_{D}+2\gamma \Gamma _{D}^{2}+(\gamma ^{2}+\Omega ^{2})\Gamma _{D}^{3}]}
\notag \\
B &=&\gamma +(\gamma ^{2}+\Omega ^{2})\Gamma _{D}, \nonumber
\hspace{0.55\textwidth} (3.20)
\end{eqnarray}%
where $g^{2}\Omega $ and $g^{2}\gamma $ are the exciton frequency shift and
decay rate, respectively. The derivation of the current equation is closely
analogous to the spontaneous emission of phonons in double dots \cite{46},
in which the correlation functions $\left\langle X_{t}X_{t^{\prime
}}^{\dagger }\right\rangle _{0}$ is given by the electron-phonon interaction.

As can be seen from Eq. (3.20), the stationary current through the quantum
dot depends strongly on the decay rate $\gamma $. The tunnel currents of a
quantum dot inside a planar microcavity is numerically displayed in Fig. 5.
In plotting the figure, the current is in terms of 100 pA, and the cavity
length is in units of $\lambda _{0}/2$, where $\lambda _{0}$ is the
wavelength of the emitted photon. Furthermore, the tunneling rates, $\Gamma
_{U}$ and $\Gamma _{D}$, are assumed to be equal to 0.2$\gamma _{0}$ and $%
\gamma _{0},$ respectively. Here, a value of 1/1.3ns for the free-space
quantum dot decay rate $\gamma _{0}$ is used in our calculations \cite{47}.
Also, the planar microcavity has a Lorentzian broadening at each resonant
modes (with broadening width equals to 1\% of each resonant mode) \cite{38}.
As the cavity length is less than half of the wavelength of the emitted
photon, the stationary current is inhibited. This is because the energy of
the photon generated by the quantum dot is less than the cut-off frequency
of the planar microcavity. Moreover, the current is increased whenever the
cavity length is equal to multiple half wavelength of the emitted photon. It
represents as the cavity length exceeds some multiple wavelength, it opens
up another decay channel abruptly for the quantum dot exciton, and turns out
that the current is increased. With the increasing of cavity length, the
stationary current becomes less affected by the cavity and gradually
approaches to free space limit.

\begin{figure}[th]
\center\includegraphics[width=8cm,clip=true]{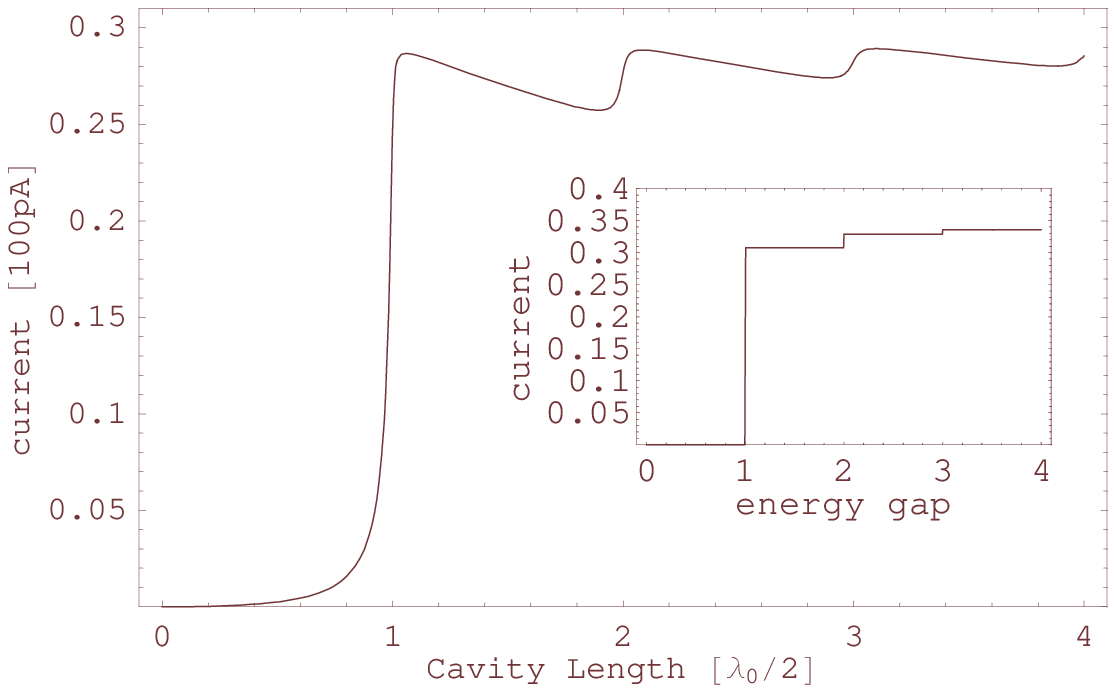}
\caption{Stationary tunnel current, Eq. (3.20), as a function of
cavity
length $L_{c}$. The vertical and horizontal units are 100 pA and $\protect%
\lambda _{0}$, respectively. Inset : $\protect\overset{\wedge }{\left\langle
I\right\rangle }$ as a function of exciton energy gap $\protect\varepsilon $%
. The cavity length is fixed to $\protect\lambda _{0}/2$. The current is in
units of 100 pA, while the energy gap is terms of $2hc/\protect\lambda _{0}$%
. }
\end{figure}

To understand the inhibited current thoroughly, we now fix the cavity length
equal to $\lambda _{0}/2$ and vary the exciton energy gap, while the planar
microcavity is now assumed to be perfect. The vertical and horizontal units
in the inset of Fig. 5 are 100 pA and $2hc/\lambda _{0}$, respectively.
Here, $\lambda _{0}$ is the wavelength of the photon emitted by the quantum
dot exciton in free space. Once again, we observe the suppressed current as
the exciton energy gap is tuned below the cut-off frequency. The plateau
features in the inset of Fig. 5 also comes from the abruptly opened decay
channels for the quantum dot exciton. From the experimental point of view,
it is not possible to tune either the cavity length or the energy gap for
such a wide range. A possible way is to vary the exciton gap around the
first discontinuous point $2hc/\lambda _{0}$. Since the discontinuities
should smear out for the real microcavity, it is likely to have a peak if
one measures the differential conductance $d\overset{\wedge }{\left\langle
I\right\rangle }/d\varepsilon $ as a function of energy gap $\varepsilon $.

\section{Current through the double-dot system and the induced entanglement}

Now, we consider \emph{two spatially separated} quantum dots incorporated
inside the \textit{p-i-n} junction. The novel feature here is the
dissipative creation of entanglement over relatively large distances, and
its readout via the stationary current. The device structure is shown in
Fig. 6.
\begin{figure}[th]
\center\includegraphics[width=8cm]{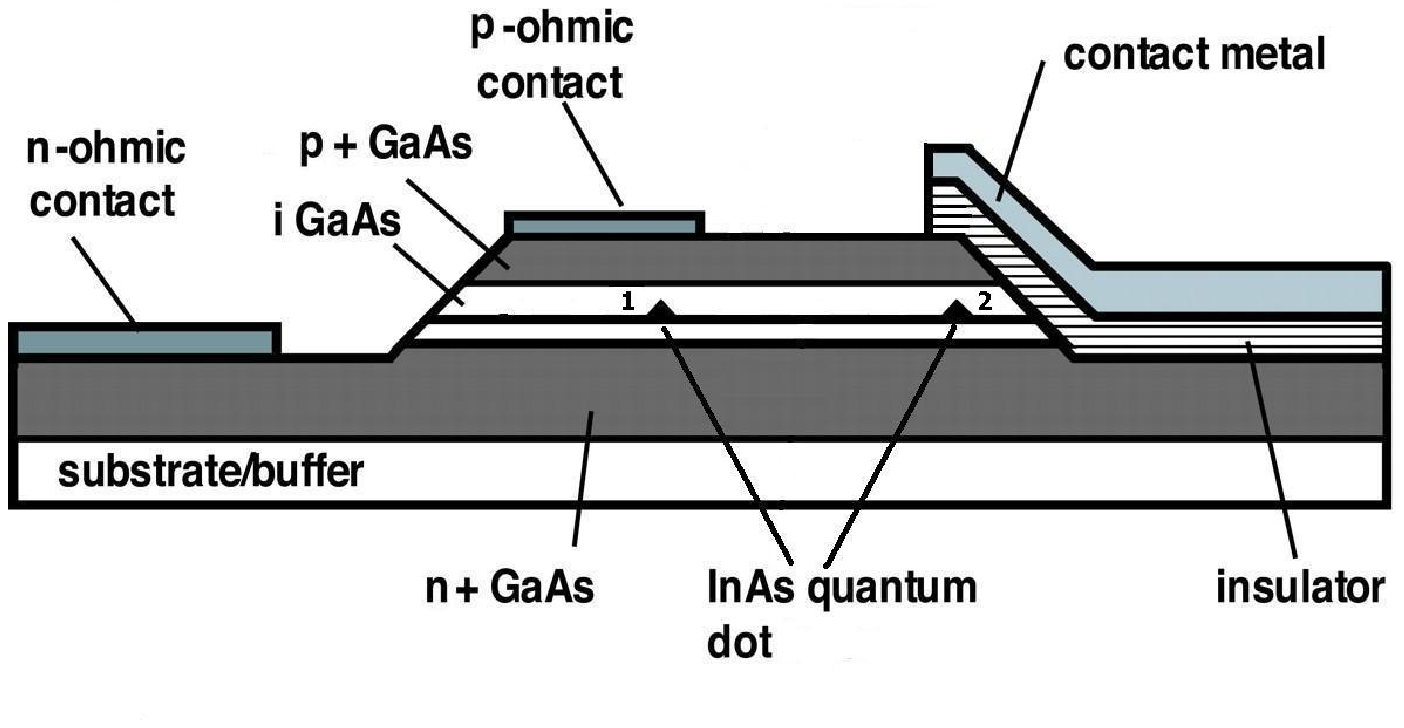}
\caption{Proposed device structure. Two InAs quantum dots are embedded in a%
\textit{\ p-i-n }junction. Above dot 2 is a metal gate, which control the
energy gap and orientation of the dipole.}
\end{figure}

One of the obstacles in measuring superradiance between the quantum dots
comes from the random size of the dots which result in a random distribution
of energy gap and thus diminishes the coherent radiation. This can be
overcome by constructing a gate voltage over one of the quantum dots. The
energy gap and the orientation of the dipole moments of one of the quantum
dots, thus can be controlled well.

After a hole is injected into the hole subband in the quantum dot, the
\textit{n}-side electron can tunnel into the exciton level because of the
Coulomb interaction between the electron and hole. In our calculation, we
also neglect the Forster process which may have some influences on the
results if the two dots are close to each other. The validity of this
assumption will be discussed later. Thus, we may assume four dot states $%
\left| 0\right\rangle =\left| 0,h;0,0\right\rangle $, $\left|
U_{1}\right\rangle =\left| e,h;0,0\right\rangle $, $\left|
U_{2}\right\rangle =\left| 0,0;e,h\right\rangle $, and $\left|
D\right\rangle =\left| 0,0;0,0\right\rangle $, where $\left|
0,h;0,0\right\rangle $ means there is one hole in dot 1 and $\left|
0,0;0,0\right\rangle $ represents the ground state with no hole and electron
in the quantum dots. The exciton states $\left| e,h;0,0\right\rangle $ (in
dot 1) can be converted to $\left| 0,0;e,h\right\rangle $ (in dot 2) through
the exciton-photon interactions. By transforming $\left| U_{1}\right\rangle $
and $\left| U_{2}\right\rangle $ into Dicke states: $\left|
S_{0}\right\rangle =\frac{1}{\sqrt{2}}\left( \left| U_{1}\right\rangle
-\left| U_{2}\right\rangle \right) $ and $\left| T_{0}\right\rangle =\frac{1%
}{\sqrt{2}}\left( \left| U_{1}\right\rangle +\left| U_{2}\right\rangle
\right) $, we can now define the dot-operators $\overset{\wedge }{n_{S}}%
\equiv \left| S_{0}\right\rangle \left\langle S_{0}\right| ,\overset{\wedge }%
{n_{T}}\equiv \left| T_{0}\right\rangle \left\langle T_{0}\right| ,$ $%
\overset{\wedge }{n_{D}}\equiv \left| D\right\rangle \left\langle D\right| ,$
$\overset{\wedge }{p_{s}}\equiv \left| S_{0}\right\rangle \left\langle
D\right| ,$ $\overset{\wedge }{p_{T}}\equiv \left| T_{0}\right\rangle
\left\langle D\right| ,\overset{\wedge }{s_{U_{1}}}\equiv \frac{1}{\sqrt{2}}%
(\left| 0\right\rangle \left\langle S_{0}\right| +\left| 0\right\rangle
\left\langle T_{0}\right| ),$ $\overset{\wedge }{s_{D}}\equiv \left|
0\right\rangle \left\langle D\right| $. The exciton-photon coupling is
described by an interaction Hamiltonian $H_{T}$:

\begin{eqnarray}
H_{T} &=&\sum_{\mathbf{k}}\frac{1}{\sqrt{2}}g\{D_{\mathbf{k}}b_{\mathbf{k}}[%
\overset{\wedge }{p_{S}}(1+e^{i\mathbf{k}\cdot \mathbf{r}})  \notag \\
&&+\overset{\wedge }{p_{T}}(1-e^{i\mathbf{k}\cdot \mathbf{r}})]+c.c.\}
\notag \\
&=&g(\overset{\wedge }{p_{S}}X_{S}+\overset{\wedge }{p_{S}}^{\dagger
}X_{S}^{\dagger }+\overset{\wedge }{p_{T}}\overline{X_{T}}+\overset{\wedge }{%
p_{T}}^{\dagger }\overline{X_{T}}^{\dagger }), \nonumber
\hspace{0.35\textwidth} (4.1)
\end{eqnarray}%
where $\mathbf{r}$ is the position vector between two quantum dot, $%
X_{S}=\sum_{\mathbf{k}}(1+e^{i\mathbf{k}\cdot \mathbf{r}})D_{\mathbf{k}}b_{%
\mathbf{k}}$, and
$\overline{X_{T}}=\sum_{\mathbf{k}}(1-e^{i\mathbf{k}\cdot
\mathbf{r}})D_{\mathbf{k}}b_{\mathbf{k}}$. The dipole approximation
is not used in our calculation since we keep the full $e^{i\mathbf{k}\cdot \mathbf{r%
}}$ terms in the operators $X_{S}$ and $\overline{X_{T}}$. Following the
derivations in previous section, one can also derive a master equation for
this double dot system. The equations of motion can be expressed as

\begin{eqnarray}
\overset{\wedge }{\left\langle n_{\sigma }\right\rangle }_{t}-\overset{%
\wedge }{\left\langle n_{\sigma }\right\rangle }_{0}
&=&-ig\int_{0}^{t}dt^{\prime }\{\overset{\wedge }{\left\langle
p_{\sigma }\right\rangle }_{t^{\prime }}-\overset{\wedge
}{\left\langle p_{\sigma }^{\dagger }\right\rangle }_{t^{\prime
}}\} \notag \\
&&+\Gamma _{U}\int_{0}^{t}dt^{\prime }(1-\overset{\wedge
}{\left\langle n_{S}\right\rangle }_{t^{\prime }}-\overset{\wedge
}{\left\langle n_{T}\right\rangle }_{t^{\prime }}-\overset{\wedge
}{\left\langle
n_{D}\right\rangle }_{t^{\prime }})  \notag \\
\overset{\wedge }{\left\langle n_{D}\right\rangle }_{t}-\overset{\wedge }{%
\left\langle n_{D}\right\rangle }_{0} &=&-ig\int_{0}^{t}dt^{\prime }\{%
\overset{\wedge }{\left\langle p_{S}\right\rangle }_{t^{\prime }}-\overset{%
\wedge }{\left\langle p_{S}^{\dagger }\right\rangle }_{t^{\prime }}+\overset{%
\wedge }{\left\langle p_{T}\right\rangle }_{t^{\prime }}-\overset{\wedge }{%
\left\langle p_{T}^{\dagger }\right\rangle }_{t^{\prime }}\} \notag
\\
&&-2\Gamma _{D}\int_{0}^{t}dt^{\prime }\overset{\wedge
}{\left\langle
n_{D}\right\rangle }_{t^{\prime }}  \notag \\
\overset{\wedge }{\left\langle p_{S}\right\rangle }_{t}-\overset{\wedge }{%
\left\langle p_{S}\right\rangle _{t}^{0}} &=&-\Gamma
_{D}\int_{0}^{t}dt^{\prime }e^{i\varepsilon (t-t^{\prime })}\left\langle
X_{t}X_{t^{\prime }}^{\dagger }\widetilde{p_{S}}(t^{\prime })\right\rangle
_{t^{\prime }}  \notag \\
&&-ig\int_{0}^{t}dt^{\prime }e^{i\varepsilon (t-t^{\prime })}\{\left\langle
\overset{\wedge }{n_{S}}X_{t}X_{t^{\prime }}^{\dagger }\right\rangle
_{t^{\prime }}-\left\langle \overset{\wedge }{n_{D}}X_{t^{\prime }}^{\dagger
}X_{t}\right\rangle _{t^{\prime }}\}  \notag \\
\overset{\wedge }{\left\langle p_{T}\right\rangle }_{t}-\overset{\wedge }{%
\left\langle p_{T}\right\rangle _{t}^{0}} &=&-\Gamma
_{D}\int_{0}^{t}dt^{\prime }e^{i\varepsilon (t-t^{\prime })}\left\langle
\overline{X}_{t}\overline{X}_{t^{\prime }}^{\dagger }\widetilde{p_{T}}%
(t^{\prime })\right\rangle _{t^{\prime }}  \notag \\
&&-ig\int_{0}^{t}dt^{\prime }e^{i\varepsilon (t-t^{\prime })}\{\left\langle
\overset{\wedge }{n_{T}}\overline{X}_{t}\overline{X}_{t^{\prime }}^{\dagger
}\right\rangle _{t^{\prime }}-\left\langle \overset{\wedge }{n_{D}}\overline{%
X}_{t^{\prime }}^{\dagger }\overline{X}_{t}\right\rangle _{t^{\prime
}}\}, \nonumber \hspace{0.10\textwidth} (4.2)
\end{eqnarray}%
where the index $\sigma =S$ or $T$.

Similarly, the tunnel current $\widehat{I}$ can be defined as the change of
the occupation of $\overset{\wedge }{n_{D}}$ and is given by $\widehat{I}%
\equiv ig\sum_{\sigma }(\overset{\wedge }{p_{\sigma }}-\overset{\wedge }{%
p_{\sigma }}^{\dagger })$. The expectation value $\overset{\wedge }{%
\left\langle I\right\rangle }_{t}$ can be obtained in the limit of $%
t\rightarrow \infty $ and reads

\begin{equation}
\overset{\wedge }{\left\langle I\right\rangle }_{t\rightarrow \infty }=\frac{%
4g^{2}\gamma _{+}\gamma _{-}}{\gamma _{-}+\gamma _{+}[1+2\gamma
_{-}(g^{2}/\Gamma _{D}+g^{2}/\Gamma _{U}+\Gamma _{D})]},  \tag{4.3}
\end{equation}%
where $g^{2}\gamma _{+}$ and $g^{2}\gamma _{-}$ are the superradiant and
subradiant decay rate of the exciton, respectively.

The corresponding decay rate for superradiant and the subradiant channels is
given by

\begin{equation}
g^{2}\gamma _{\pm }=\gamma _{0}(1\pm \frac{\sin (2\pi d/\lambda _{0})}{2\pi
d/\lambda _{0}}),  \tag{4.4}
\end{equation}%
where $d$ is the inter-dot distance and $\gamma _{0}$ is the exciton decay
rate in a quantum dot. To display the dependence of the stationary current
through the quantum dot on the dot distance $d$, we present the results of
two identical quantum dots in Fig. 7.
\begin{figure}[h]
\center\includegraphics[width=7.5cm]{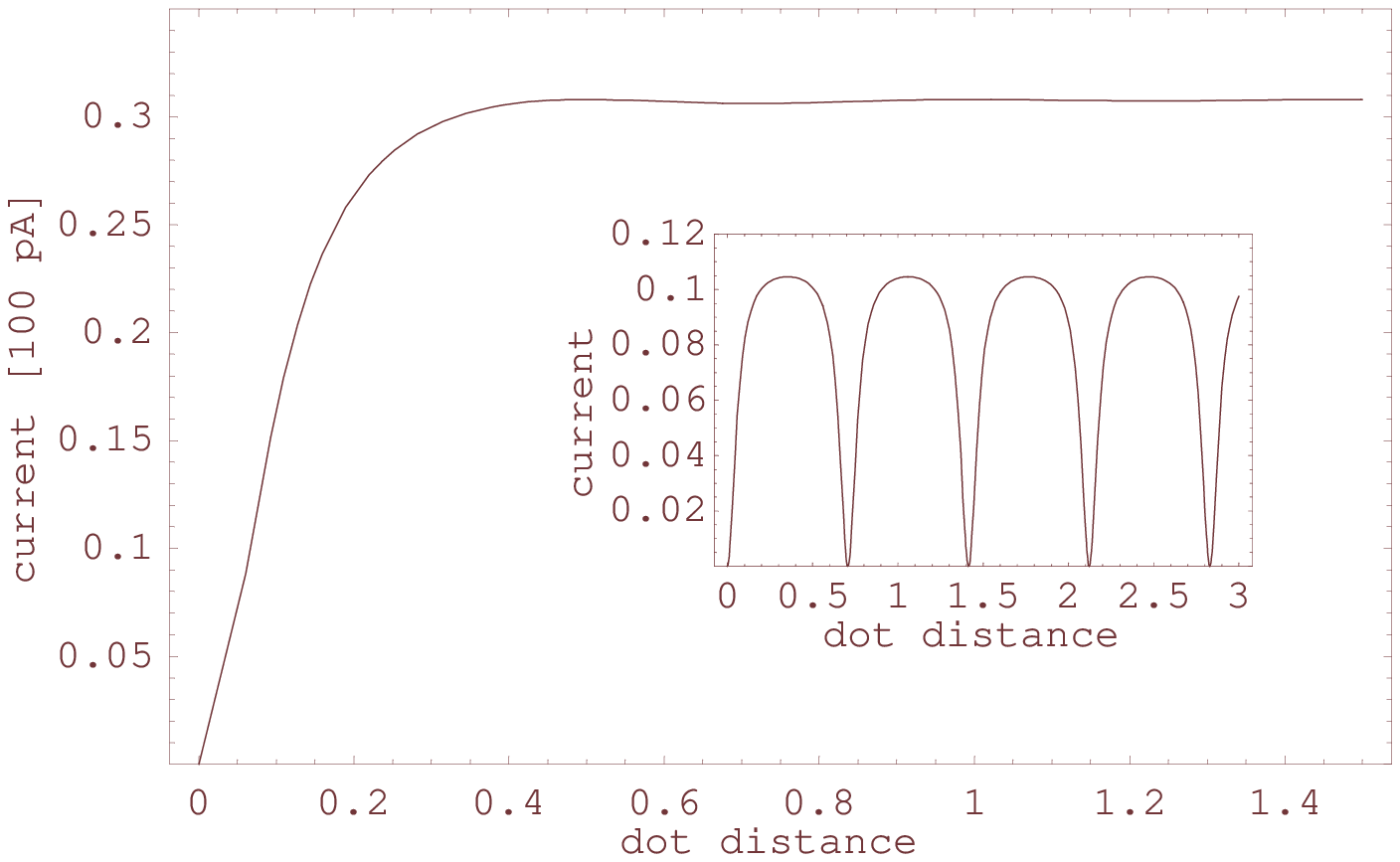} \caption{{}Stationary
tunnel current as a function of dot distance $d$. The interference
effect is seen clearly (inset) by incorporating the system inside a
rectangular microcavity. The vertical and horizontal units are 100
pA and $\protect\lambda _{0}$, respectively.}
\end{figure}
As shown in Fig. 7, the current is suppressed as the dot distance $d$ is
much smaller than the wavelength ($\lambda _{0}$) of the emitted photon.
This corresponds to the trapping state in the two-ion system. As long as we
choose only one of the dots to be coupled to reservoirs, the generated
photon is reabsorbed immediately by the other dot and vice versa. The
current is then blocked by this exchange process. For small rates limit ($%
g^{2}\gamma _{\pm }$) one can approximate Eq. (4.3) by $I\approx
4[1/g^{2}\gamma _{-}+1/g^{2}\gamma _{+}]^{-1}$. The rates $\Gamma _{U,D}$
drop out completely and the current is only determined by the (smaller)
radiative decay rates. In this approximate form, the current looks identical
to the expression for the conductance $G\propto \lbrack 1/\Gamma
_{L}+1/\Gamma _{R}]^{-1}$ from a left lead through a single level to a right
lead with tunnel rates $\Gamma _{L,R}$. This implies that the superradiant
and the subradiant channel are in series (and not in parallel) in this
limit. This is because once the exciton is formed in dot 1, time evolution
of this state is proportional to $e^{-g^{2}\gamma _{+}t}+e^{-g^{2}\gamma
_{-}t}$ not $e^{-g^{2}(\gamma _{+}+\gamma _{-})t}$\cite{28}. It means the
two decay channels in our system are not in parallel. For long time behavior
$t\rightarrow \infty $ and $\gamma _{+}>>\gamma _{-}$, the function $%
e^{-g^{2}\gamma _{+}t}+e^{-g^{2}\gamma _{-}t}$ approaches the limit of $%
e^{-g^{2}\gamma _{-}t}$, which is identical to the same limit of the
function $e^{-\frac{g^{2}\gamma _{+}\gamma _{-}}{\gamma _{+}+\gamma _{-}}t}$%
(in series).

Similar to the two-ion superradiance \cite{42}, the current also exhibits
oscillatory behavior as a function of dot distance. To observe the
interference effect clearly, one may incorporate the system inside a
microcavity since semiconductor cavities with strong electron-photon
coupling have been realized experimentally by, e.g., G\'{e}rard\textit{\ et
al}.\cite{48}. Reduction of the allowed $k$-state is expected to increase
the magnitude of the oscillation. For example, if the system is placed
inside a rectangular microcavity with length $\lambda _{0}$, the decay rate
for the two channels can be worked out straightforwardly:

\begin{equation}
g^{2}\gamma _{cav,\pm }=\frac{\gamma _{0}}{\pi }\left| 1\pm e^{i2\pi d/(%
\sqrt{2}\lambda _{0})}\right| ^{2}.  \tag{4.5}
\end{equation}%
The stationary current is plotted in the inset of Fig. 7, where a perfect
(lossless) cavity is assumed. As we mentioned above, the amplitude of
oscillation is larger than that in free space. However, the oscillation
period is not half of the wavelength, but $\lambda _{0}/\sqrt{2}$. This is
because the interference term is only influenced by the wave vector in the
unconfined direction. Excluding the contributions from fundamental cavity
modes, the effective wave vector can be expressed as

\begin{equation}
k_{eff}=\sqrt{(\frac{2\pi }{\lambda _{0}})^{2}-2\ast (\frac{\pi }{\lambda
_{0}})^{2}}=\frac{k_{0}}{\sqrt{2}}.  \tag{4.6}
\end{equation}%
The oscillation period of the decay rate and the current is therefore
increased by a factor of $\sqrt{2}$.

In Fig. 8, we plot the expectation value of $n_{S\text{ }}$ ($n_{T}$) as a
function of the dot distance. The maximum entangled state ($\left|
S_{0}\right\rangle $) is reached as $d<<$ $\lambda _{0}$. This is remarkable
as the steady state is independent of the initial state. The entanglement is
induced by the cooperative decoherence in the system. In a recent paper by
Schnider \textit{et al}.\cite{49}, the authors consider the behavior of an
ion trap with all ions driven simultaneously and coupled collectively to a
heat bath. They also found that the steady state of the ion trap can exhibit
quantum entanglement. However, the concurrence of their system is below the
value of unity (maximum entanglement). On the contrary, in our system the
maximum entangled state can be generated by tuning the band gap of dot 2
(linear stark effect), i.e. control the on/off of the superradiance. Another
advantage of our scheme is shown in the inset of Fig. 8. If the double-dot
system is incorporated inside a rectangular microcavity, the maximum
entangled states repeat as a function of inter-dot distance. This means even
for remote separation, the entanglement can still be achieved. The reason
can be attributed to that the creation of entanglement in our model is
governed by the interaction with a common heat bath \cite{50}, while
conventional creation of entanglement depends on the direct interaction
between two subsystems\cite{51}. When two dots are coupled to the common
photon fields, the collective decay process drives the system into the
entangled states. The novel feature of the effect predicted here is that
entanglement in fact can be controlled electrically (without applying a
laser field) and read out in the form of a transport property, i.e., the
electron \emph{current} (as a function of the dot distance or,
alternatively, the cavity length).
\begin{figure}[h]
\center\includegraphics[width=7.5cm]{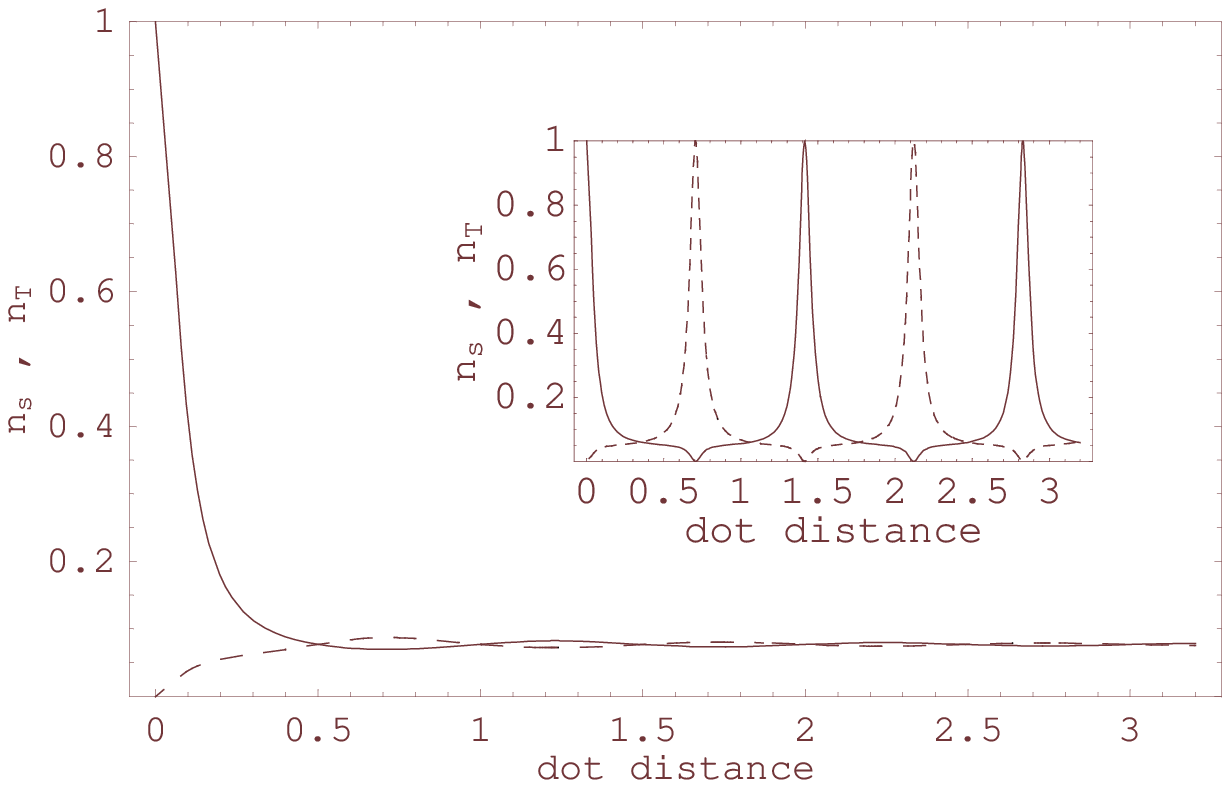} \caption{{}Occupation
probability of the entangled states $n_{S\text{ }}$ (solid line) and
$n_{T}$ (dashed line). The inset shows the results inside a
rectangular microcavity.}
\end{figure}

Another possible application of this effect is that by tuning the coherence
of the dots, one can control the emission of single photon at predetermined
times, which is important for the field of quantum information technology.
One might argue that for small inter-dot distance the Forster process may
play some role in our system\cite{52}; nevertheless, this only causes small
energy splitting between state $\left| S_{0}\right\rangle $ and $\left|
T_{0}\right\rangle $. Comparing to the large energy difference in the III-V
semiconductor material, its effect on the decay rate $g^{2}\gamma _{\pm }$
is negligible.

A few remarks about the problem of dissipation should be mentioned here. The
coherence of the quantum states is a fundamental issue in quantum physics
and decoherence caused by phonons or imperfections may destroy the unitary
quantum evolution. In our proposals, decoherence due to interaction with
other bosonic excitations (phonons and electron-hole pairs in the leads) is
inevitable but can in principle be (partly) controlled by variation of the
dot energies, or control of the mechanical degree of freedom\cite{53}. In
addition, scattering due to impurities are negligible since there is no
interdot transport in our system.

\section{Conclusions}

In summary, we have proposed a new method of detecting superradiant and
Purcell effects in semiconductor quantum dots. By incorporating a quantum
dot between a p-i-n junction, the Purcell effect on stationary tunnel
current can be examined by the variations of cavity length or exciton gap.
For the double-dot system, the superradiant effect on the stationary tunnel
current can be examined by tuning the band gap of the quantum dot. The
interference effects between two dots can be seen more clearly by
incorporating the system inside a microcavity. The oscillation period of the
decay rate and current is also increased because of the microcavity.
Moreover, the maximum entangled state is induced as the inter-dot distance
is much smaller than the wavelength of the emitted photon. Our model
provides a new way to generate the entanglement in solid-state systems and
maybe useful in future quantum information processing.

\section{Acknowledgments}
This work is supported partially by the National Science Council,
Taiwan under the grant number NSC 93-2112-M-009-037.

\end{document}